\documentclass[iop,apj,onecolumn]{emulateapj}
\slugcomment{Accepted for publication in ApJ}
\usepackage{xspace}
\usepackage{apjfonts}
\usepackage{color}
\shorttitle{Cosmic-Rays and Cool Clouds}
\shortauthors{Everett \& Zweibel}

\newcommand{\muG}{{\,$\mu$\rm G}\xspace}

\newcommand{\percc}{{\rm\,cm$^{-3}$}\xspace}

\newcommand{\pc}{{\rm\,pc}\xspace}

\newcommand{\MeV}{{\rm\,Me\hspace{-.12ex}V}\xspace}
\newcommand{\GeV}{{\rm\,Ge\hspace{-.12ex}V}\xspace}

\newcommand{\alfven}{{Alfv\'en}\xspace}
\newcommand{\vA}{v_{\hspace{-.20ex}\rm A}}
\newcommand{\bfvA}{\bf v_{\hspace{-.20ex}\rm A}}
\newcommand{\Pcr}{P_{\rm cr}\xspace}
\newcommand{\Pw}{P_{\rm w}\xspace}

\begin{document}

\title{The Interaction of Cosmic Rays with Diffuse Clouds}

\author{John E. Everett and Ellen G. Zweibel}
\affil{University of Wisconsin--Madison, Department of Astronomy,\\
  University of Wisconsin--Madison, Department of Physics, \\
and Center for Magnetic Self-Organization in
Laboratory and Astrophysical Plasmas} 
\email{everett@physics.wisc.edu, zweibel@astro.wisc.edu}

\begin{abstract}
  We study the change in cosmic-ray pressure, the change in cosmic-ray
  density, and the level of cosmic-ray induced heating via
  \alfven-wave damping when cosmic rays move from a hot ionized plasma
  to a cool cloud embedded in that plasma.  The general analysis
  method outlined here can apply to diffuse clouds in either the
  ionized interstellar medium or in galactic winds.  We introduce a
  general-purpose model of cosmic-ray diffusion building upon the
  hydrodynamic approximation for cosmic rays (from McKenzie \& V\"olk
  and Breitschwerdt and collaborators).  Our improved method
  self-consistently derives the cosmic-ray flux and diffusivity under
  the assumption that the streaming instability is the dominant
  mechanism for setting the cosmic-ray flux and diffusion.  We find
  that, as expected, cosmic rays do not couple to gas within cool
  clouds (cosmic rays exert no forces inside of cool clouds), that the
  cosmic-ray density does not increase within clouds (it may slightly
  decrease in general, and decrease by an order of magnitude in some
  cases), and that cosmic-ray heating (via \alfven-wave damping and
  not collisional effects as for $\sim 10$\,\MeV cosmic rays) is only
  important under the conditions of relatively strong (10\,\muG)
  magnetic fields or high cosmic-ray pressure ($\sim
  10^{-11}$\,ergs\,cm$^{-3}$).
\end{abstract}

\section{Introduction}\label{Intro}

The possibility of a rise or fall in cosmic-ray density within diffuse
clouds and/or molecular cores has been studied over many years by
researchers interested either in gamma-ray emission from cool clouds
\citep[e.g.,][]{Skilling1971, SkillingStrong1976, CesarskyVoelk1977,
  StrongSkilling1977, CesarskyVoelk1978, Morfill1982a, Morfill1982b,
  Voelk1983} or in ionization by cosmic rays within clouds
\citep[e.g.,][]{HartquistEtAl1978, SuchkovEtAl1993, PadoanScalo2005,
  PadovaniEtAl2009, Papadopoulos2010}.  Another regime where this
interaction may be important is in cool clouds that are readily
observed within galactic winds \citep[e.g.,][]{HeckmanEtAl1990,
  Martin2005, RupkeEtAl2005, SteidelEtAl2010}.  In the context of such
winds, cosmic rays are known to help heat and drive outflows of fully
ionized gas \citep{BreitschwerdtEtAl1991, EverettEtAl2008,
  SocratesEtAl2008, EverettEtAl2010}; can they help heat and
accelerate cool clouds embedded in a hot wind?

We will examine this problem using a set of hydrodynamic equations
that treat both convective and diffusive transport of cosmic rays
self-consistently; this has rarely been done in studies of cosmic-ray
dynamics, and has never been done in the context of cool clouds.
Before we start, however, it is important to understand the variety of
previous work on this problem.  We start, therefore, with a review of
models of cosmic-ray interaction with cool clouds in
Section~\ref{reviewSection}; we cover models of cosmic-rays in the ISM
(Section~\ref{ismStudiesReview}) \& cosmic rays in galactic winds
(Section~\ref{galacticWindsReview}), followed by an introduction to
our method (Section~\ref{ourModel}).  Section~\ref{cosmicRayHydro}
defines our equations for cosmic-ray hydrodynamics, and
Section~\ref{cloudModels} applies those models to a simple
hot-gas/cool-cloud interface, and examines how robust those results
are.  Finally, Section~\ref{conclusions} summarizes our results and
compares \& contrasts those results with previous work.

\section{Review of Previous Studies of Cosmic Rays in Inhomogeneous Media}\label{reviewSection}

\subsection{Cosmic Rays in the Multiphase Interstellar Medium}\label{ismStudiesReview}

Galactic cosmic rays are thought to be energized in shocks, most
likely in supernova remnants, but perhaps in other shocks as well
\citep[see, e.g.,][]{BerezinskiiEtAl1990, Schlickeiser2002,
  Hillas2006}.  Cosmic rays then propagate outward along magnetic
field lines, moving through the interstellar medium (ISM).  Cosmic
rays gyrate around these field lines, but also change direction,
scattering off of already-existing magnetic-field inhomogeneities
\citep[such as magnetic mirrors,
e.g.,][]{CesarskyVoelk1978,Chandran2000a}, ambient turbulence
\citep[e.g.,][]{Jokipii1966, Chandran2000b,YanLazarian2004}, and
\alfven waves generated resonantly by the cosmic rays themselves
\citep[the streaming instability; see][]{Lerche1966, Lerche1967,
  Wentzel1968, KulsrudPearce1969, Tademaru1969, Wentzel1969,
  Skilling1970, Skilling1971, Skilling1975, Kulsrud2005}.  In this
work, and in the review in this section, we will concentrate on
cosmic-ray scattering due to turbulence generated by cosmic rays
themselves.  Not only is the streaming instability a leading candidate
for the propagation of $\sim 1-100$\,\GeV cosmic rays, but focusing
on the streaming instability helps to demarcate a relatively
well-defined problem where we can self-consistently define a
cosmic-ray flux and diffusivity.  Also, the streaming instability is a
key component of any study of the interaction of cosmic rays with
magnetic fields in the hot ISM \citep[see also][who showed that no
other turbulent-wave energy source would lead to significant
scattering within clouds]{CesarskyVoelk1978}.

First, we give a short overview of the streaming instability
\citep[see also][]{Lerche1966, Wentzel1968, KulsrudPearce1969,
  Tademaru1969, Wentzel1969, Skilling1971, Zweibel2003, Kulsrud2005,
  EverettEtAl2008}; understanding this instability will be essential
in understanding previous results. Simply put, the streaming
instability feeds off of the drifting motion of the cosmic-ray
population: for instance, the cosmic rays may drift as they stream
down their own density gradient.  The streaming instability is
launched if they move (in bulk) more quickly than the local \alfven
speed; in this case, their cyclotron motion around magnetic-field
lines excites \alfven waves with wavelength of order the cosmic-ray
gyroradius, as they scatter off of \alfven waves.  If those waves are
not damped much more quickly than they are excited, the \alfven waves
will then scatter the cosmic rays, limiting their bulk motion to
approximately the local \alfven speed.  In applying this mechanism to
the ionized ISM, the amplitude of the \alfven waves can be quite small
($\delta B \sim 10^{-3}B_{\rm ambient}$); with an \alfven wave of this
strength, the cosmic-ray mean-free path is $\lambda_{\rm mfp} \sim
0.1$\,pc \citep[see][]{Kulsrud2005}.  The excitation time-scale for
these waves is of order $10^4$\,years in the ISM, so as long as the
damping rates are not faster than this, this instability will be
important.  Of course, as we show in this paper, as the balance
between wave damping and wave growth changes, the cosmic-ray mean-free
path varies strongly from $\la 0.01$\,pc to larger than the size of
diffuse clouds; this will be shown most explicitly by the variation in
$\kappa_{\rm cr}$, explored in Section~\ref{pressureChanges}.

The study of the streaming instability and cosmic-ray propagation from
ionized plasma into diffuse clouds started with
\citet{KulsrudPearce1969}, who pointed out the importance of
ion-neutral damping to quickly destroy \alfven waves generated by the
streaming instability \citep[see also][]{KulsrudCesarsky1971}.  They
inferred from this that one could ignore the interstellar clouds for
the purposes of cosmic-ray diffusion: in their work, cosmic rays move
at approximately the \alfven speed in the ionized regions of the ISM
only, and free-stream at $v \sim c$ in molecular clouds.  A similar
result was obtained by \citet{Skilling1971}, who named such regions
``free zones'': these are regions where cosmic rays are not locked to
the \alfven speed because of wave damping

After these initial studies, interest in gamma-ray emission from
clouds brought this problem back to prominence, with
\citet{SkillingStrong1976} and \citet{CesarskyVoelk1977} further
investigating how cosmic rays interact with clouds \citep[see
also][]{CesarskyEtAl1977, StrongSkilling1977, CesarskyVoelk1978,
  HartmanEtAl1979, Voelk1983}.  \citet{SkillingStrong1976} realized
that the destruction of low-energy ($E \la 300$\,\MeV) cosmic-ray
protons within the cloud, due to ionization and nuclear interactions,
causes a decrease in the cosmic-ray density at these low energies.
This means that the cosmic-ray flux exiting a cloud is lower than the
input cosmic-ray flux, and this flux difference excites the streaming
instability: cosmic-ray generated \alfven waves grow in a cosmic-ray
density gradient.  The streaming instability then generates \alfven
waves, which act as scattering centers for these low-energy cosmic
rays, slowing down these cosmic rays from their free-streaming
velocity to approximately the \alfven speed, $\vA$.  This decreases
the rate at which cosmic rays can `flow' into the cloud, or put
another way, it slows the rate at which cosmic rays are replenished
within the cloud; \citet{SkillingStrong1976} referred to this process
as therefore ``excluding'' cosmic rays from the cloud.
\citet{CesarskyVoelk1977} studied this mechanism (as well as a variety
of others), finding that this exclusion would occur only for cosmic
rays with $E \la 50$\,\MeV if the magnetic field strength in the cloud
increased to 50\muG as opposed to the constant 3\muG assumed in
\citet{SkillingStrong1976}; the compression of the magnetic field
raises the cosmic-ray flux into the cloud, and as a result, only the
lowest-energy cosmic rays (again, $E \la 50$\,\MeV) are not
replenished quickly enough in the cloud \citep{CesarskyVoelk1977,
  Voelk1983}.  (We note that very recent observations of magnetic
field strength in cool phases of the ISM seem to show that the
magnetic field does not increase significantly in strength for $n \la
300$\,cm$^{-3}$; see \citealt{CrutcherEtAl2010}.) It is important to
note that these previous analyses seems to rely upon a uniform
distribution of cosmic rays far away from the cloud, so that the
streaming instability only operates in the ionized region on the
periphery of the clouds \citep[also, the importance of the interaction
of multiple clouds is briefly addressed in][]{Morfill1982b}.

We mention briefly, for completeness, that cosmic-ray electrons could
also be impacted in these processes. \citet{CesarskyVoelk1978} pointed
out that scattering of cosmic-ray electrons off of the \alfven waves
generated by cosmic-ray protons with the same gyro-radii would help
slow cosmic-ray electrons within clouds.  \citet{Morfill1982a} and
\citet{Morfill1982b} expanded on this idea by considering how
cosmic-ray electrons might be accelerated within the clouds and how
secondaries would be generated within the cloud, helping to increase
the $\gamma$-ray brehm{\ss}trahlung emission.

But again, the above mechanisms operate on cosmic rays at relatively
low energies (of order $50$\,\MeV or lower), and so would not exclude
$\sim$1\,\GeV cosmic rays from clouds.  This is important for this
study, as $\sim$1\,\GeV cosmic rays are the dominant source of
cosmic-ray pressure, which is our prime interest here.  So, while this
``exclusion'' effect may occur in the ISM, and this mechanism would be
important for understanding the ionization of cool clouds by cosmic
rays \citep[][]{HartquistEtAl1978, SuchkovEtAl1993, PadovaniEtAl2009,
  Papadopoulos2010}, it will not strongly impact cosmic-ray pressure
or \GeV gamma-ray emission.

After the above papers were published, it seems that cosmic rays were
then left to stream on their own for a few years.  However, in the
past decade, as cosmic-ray ionization rates have become better
constrained \citep[e.g.,][]{McCallEtAl2003, PadovaniEtAl2009,
  IndrioloEtAl2010b, IndrioloEtAl2010a}, and as observations have
started to focus on interactions of cosmic rays with molecular clouds
\citep[e.g.,][]{Aharonian2001, GabiciEtAl2007, ProtheroeEtAl2008,
  TorresEtAl2005, TorresEtAl2008, LiChen2010, GabiciEtAl2010,
  FatuzzoEtAl2010, CasanovaEtAl2011, UchiyamaEtAl2011, OhiraEtAl2011,
  EllisonBykov2011}, interest in understanding the interaction of
clouds and cosmic rays has increased.  However, almost all of the
above works consider parameterized cosmic-ray diffusion, in sharp
contrast to the models of the late-70s and early-80s, which worked to
understand cosmic-ray diffusion by studying the streaming instability.

An exception to this trend is the work of \citet{PadoanScalo2005}.
These authors calculated that, as cosmic rays move from the hot,
ionized medium into diffuse clouds, the cosmic-ray density strongly
increases (with $n_{\rm cr} \propto n_{\rm i}^{1/2}$ up to
approximately $n_{\rm i} \sim 100$\,cm$^{-3}$); they also derived that
$n_{\rm cr}$ is approximately constant in molecular clouds of higher
densities.  Their work is a response to the observations of
\citet{McCallEtAl2003}, who infer an increase in the
cosmic-ray-induced ionization rate with clouds to explain observations
of $H_3^+$.  The theoretical result of \citet{PadoanScalo2005} seems
to contradict, however, the above-mentioned, earlier studies which
showed an effect only for low-energy cosmic rays, and predicted only a
\textit{decrease} in cosmic-ray density at those energies.  How do we
understand this difference?  How do we predict the pressure within
clouds, or the gamma-ray flux from diffuse clouds and molecular
clouds?

\subsection{Cosmic Rays in Multiphase Galactic Winds}\label{galacticWindsReview}

Galactic winds also have multiple gas phases, just like the ISM that
launches those winds.  For instance, Na~\textsc{I} and Mg~\textsc{II}
absorbers have been detected in the halos of luminous and
ultra-luminous infrared galaxies; these neutral and near-neutral
components are apparently outflowing from their host galaxies with
velocities of hundreds of kilometers per second
\citep{HeckmanEtAl1990, HeckmanEtAl2000, ShapleyEtAl2003, Martin2005,
  RupkeEtAl2005, SteidelEtAl2010}.  The clouds \citep[observed to have
small line-of-sight covering fraction; see][]{ChenEtAl2010} are
important for a number of reasons.  First, the velocities of cool
clouds are much easier to measure than the velocity of the hot-gas
component in winds, and so such clouds can be used to constrain the
velocities of the hot wind (if we understand how the clouds interact
with the hot wind).  Second, it is hard to understand how such clouds
survive (or perhaps re-form?) as coherent entities in the wind for
timescales approaching tens of millions of years
\citep{MarcoliniEtAl2005, CooperEtAl2008, CooperEtAl2009,
  MurrayEtAl2010}.  Interestingly, to our knowledge, there has been no
consideration of how cosmic rays \citep[which could help drive a
galactic wind; see][]{Ipavich1975, BreitschwerdtEtAl1991,
  EverettEtAl2008, SocratesEtAl2008} would interact with clouds in
galactic winds.

We point out that previous work has considered other driving
mechanisms for such cool clouds: in particular, much research on this
problem has focused on the radiative driving of clouds via radiative
acceleration on dust in the clouds \citep[e.g.,][]{NathSilk2009,
MurrayEtAl2010}, or perhaps by ram pressure from the surrounding
hot-phase component of the wind \citep{MurrayEtAl2005,
MurrayEtAl2010}.  However, since cosmic rays are already important in
winds, could they also help accelerate clouds?  And,
since cosmic rays can penetrate throughout each cloud, could they help
to drive the entire mass of a cloud (instead of only pushing on the
boundaries) without introducing instabilities on the cloud boundary
that could destroy it?

\subsection{An Overview of Our Approach To Both Problems}\label{ourModel}

For this paper, we will concentrate on the general interaction of
$\sim$1\,\GeV cosmic rays with clouds; we will consider parameters
applicable both to diffuse clouds embedded in the $\sim 10^6$\,K,
ionized ISM as well as cool clouds in galactic winds.  We do not
explicitly treat the interaction with molecular clouds; however, as we
will see, cosmic rays already start to freely stream (without
scattering) in diffuse clouds and their behavior will not change
qualitatively in molecular clouds, with the exception that
  cosmic-ray losses will become more extreme at high density.  In
this sense, as long as our assumptions of, for instance,
magnetic-field geometry hold, our work can be extended to molecular
clouds as well.

We find that it is relatively straightforward to investigate these
problems (under some important restrictions) using the equations of
cosmic-ray hydrodynamics \citep{McKenzieVoelk1981, McKenzieVoelk1982,
McKenzieWebb1984, BreitschwerdtEtAl1991, ZirakashviliEtAl1996}; we
have used similar equations in previous papers \citep{EverettEtAl2008,
EverettEtAl2010}.  These equations describe the interaction of cosmic
rays with an ambient magnetized medium through the action of the
streaming instability; as long as the cosmic-ray mean-free path is
larger than the scales of interest in the problem, we can use these
equations to model the cosmic-ray pressure, \alfven wave pressure, and
the diffusivity of cosmic rays.  These equations also assume that the
cosmic-rays are scattered strongly enough that they are nearly
isotropic in pitch angle (the angle between the cosmic-ray velocity
vector and the magnetic field); for the strong scattering in the
ionized phase of the ISM, and for cosmic rays of energy near 1\,\GeV,
both of these requirements are easily satisfied.

For the case of clouds in the ISM, we will investigate the cosmic-ray
pressure and density in the cloud as a function of the boundary
conditions for those diffuse clouds.  We will ask whether there is an
appreciable cosmic-ray pressure gradient in such clouds for cosmic
rays with $E \sim 1$\,\GeV, and whether the heating due to damped
\alfven waves on the periphery of the cloud is significant.

We will ask the same questions for cosmic rays interacting with clouds
within a galactic wind: what is the cosmic-ray pressure within clouds?
If the cosmic-ray density or pressure actually increases within clouds
as postulated by \citet{PadoanScalo2005}, can cosmic rays help destroy
the cloud?  Will the cosmic rays spark significant heating within the
cloud or at least at the cloud boundary?  As mentioned above, if these
clouds are to be used to understand mass outflow from galaxies
(especially as it relates to a surrounding, hot outflowing wind
component), understanding the acceleration mechanism of these clouds,
and their interaction with all of the components of the hot wind
around them, are both very important.

Before explaining the setup of our models, we stress that, of course,
the idea of a ``cloud'' itself is an idealization, as is a particular
structure for the cloud and its interface with the ambient medium.
Given the uncertainties here, we do not intend to present a study of
the interactions of cosmic rays with any particular wind- or ISM-cloud
structure; our goal is to understand how cosmic rays interact with
generic cool clouds under a range of simplified assumptions, and to
therefore understand generically what role cosmic rays can play in
this multiphase medium, and to point the way to observational tests of
the role of cosmic rays in a variety of multiphase media.

\section{Cosmic-Ray Hydrodynamics}\label{cosmicRayHydro}

When writing the equations of cosmic-ray hydrodynamics, researchers
usually assume that there is zero cosmic-ray diffusivity with respect
to \alfven waves
\citep[e.g.,][]{BreitschwerdtEtAl1991,EverettEtAl2008}.  In the case
of cool clouds, however, it is known that ion-neutral friction in the
cool clouds will damp the cosmic-ray generated \alfven waves so
strongly \citep[see.,e.g.,][]{ZweibelShull1982, Balsara1996} that it
is possible for cosmic rays to be very loosely coupled to the \alfven
waves, and diffuse relative to the waves
\citep[e.g.,][]{KulsrudCesarsky1971}.  Therefore, for this work, we
utilize the cosmic-ray hydrodynamic equations \citep[as derived and
presented in][]{McKenzieVoelk1981, McKenzieVoelk1982,
  McKenzieWebb1984, BreitschwerdtEtAl1991, ZirakashviliEtAl1996,
  PtuskinEtAl1997}, retaining all of the usual terms in the above
papers, and also including a diffusion term, where the diffusion is
set self-consistently by the \alfven-wave pressure in the medium
\citep[see also][who included a diffusion term in their calculation of
wind dynamics]{PtuskinEtAl1997}.

In all of our equations, we use Gaussian-cgs units.

\subsection{Cosmic-Ray Hydrodynamics Equations With Diffusion}

We start with the cosmic-ray transport equation \citep[Eq.~A8
from][]{BreitschwerdtEtAl1991}:
\begin{eqnarray}
\frac{\partial}{\partial t} \left( \frac{P_{\rm cr}}{\gamma_{\rm cr} - 1} \right)
& + &
\nabla \cdot \left( \frac{\gamma_{\rm cr}}{\gamma_{\rm cr}-1} ({\bf u} + {\bf v}_{\hspace{-0.20ex}\rm A})
P_{\rm cr}   -
 \frac{\kappa_{\rm cr}}{\gamma_{\rm cr} - 1} \nabla P_{\rm cr} \right)
 =  ({\bf u} + {\bf v}_{\hspace{-0.20ex}\rm A}) \nabla P_{\rm cr} +
 Q, \label{primaryDiffusionEquation}
\end{eqnarray}
where $P_{\rm cr}$ is the cosmic-ray pressure, $\gamma_{\rm cr}$ is
the adiabatic index for cosmic rays, {\bf u} is the velocity of the
thermal plasma, $\bfvA$ is the \alfven speed, $\kappa_{\rm
cr}$ is the cosmic-ray diffusivity, and $Q$ represents energy
input/loss for the cosmic rays.  Assuming a time-steady source of
cosmic-rays, and evaluating the equation for a quasi-1D system,
Equation~\ref{primaryDiffusionEquation} becomes:
\begin{eqnarray}
  \kappa_{\rm cr} \frac{d^2P_{cr}}{dz^2} & = & \left(\frac{\kappa_{\rm cr}}{P_w}
    \frac{dP_w}{dz}  +  (u + \vA) \right) \frac{dP_{cr}}{dz}  -  \frac{\gamma_{\rm cr}
    P_{cr}}{\rho} \left( \left( u + \frac{\vA}{2} \right) - q \right)
  \frac{d\rho}{dz} 
 - (\gamma_{\rm cr} - 1) Q, \label{origCRTransportEq}
\end{eqnarray}
where $\rho$ represents the plasma mass density, and 
\begin{equation}
P_{\rm w} = \frac{\langle (\delta B)^2 \rangle}{8 \pi}\label{wavePressureEq}
\end{equation}
is the \alfven-wave pressure.  We can simplify this further by setting
$u \equiv 0$ and $q \equiv 0$ (no large-scale flows and no plasma mass
sources):
\begin{eqnarray}
  \kappa_{\rm cr} \frac{d^2P_{cr}}{dz^2} & = & \left(\frac{\kappa_{\rm cr}}{P_w}
    \frac{dP_w}{dz} + \vA \right) \frac{dP_{cr}}{dz} - \left(
    \frac{\gamma_{\rm cr}
      P_{cr}}{\rho} \frac{\vA}{2} \right) \frac{d\rho}{dz} -  (\gamma_{\rm cr}-1)Q. \label{crPressureEq}
\end{eqnarray}
To derive the above result, we have assumed mass
conservation and magnetic-flux conservation.  In addition, we have
defined the cosmic-ray diffusivity $\kappa_{\rm cr}$ as
\begin{equation}
\kappa_{\rm cr} = v \lambda_{mfp} = \frac{4}{\pi} \frac{c}{3}
\frac{r_L}{(\delta B/B)^2},
\end{equation}
or, given the definition of the wave pressure,
Equation~\ref{wavePressureEq}, and defining the magnetic pressure as
\begin{equation}
P_{B} = \frac{B^2}{8 \pi}\label{BPressureEq},
\end{equation}
we can rewrite $\kappa_{\rm cr}$ as:
\begin{equation}
\kappa_{\rm cr} = v \lambda_{mfp} = \frac{4}{\pi} \frac{c}{3}
\frac{r_L}{P_{\rm w}/P_{B}} \label{kappaCRDefinition}.
\end{equation}
As a check, in the limit of $\kappa \rightarrow 0$, we
find
\begin{eqnarray}
\frac{dP_{cr}}{dz} = \frac{\gamma_{\rm cr} P_{\rm cr}}{\vA} \left(
\frac{\vA}{2 \rho} \frac{d\rho}{dz}\right)+
\frac{(\gamma_{\rm cr} - 1) Q}{\vA}\label{crPressureEqSimplified},
\end{eqnarray}
which, if we further assume $Q = 0$, simplifies to:
\begin{equation}
\frac{dP_{cr}}{dz} = \frac{\gamma_{\rm cr} P_{\rm cr}}{2 \rho} \frac{d
\rho}{dz}.
\end{equation}
This equation leads to the useful result, \textit{for cases without
diffusion},
\begin{equation}
P_{cr}\vA^{\gamma_{\rm cr}} = {\rm constant},\label{adiabaticConstant}
\end{equation}
or, taking into account that $\nabla \cdot B = 0$, we can rewrite this
as
\begin{equation}
P_{\rm cr} \rho^{-\gamma_{\rm cr}/2} = {\rm constant},
\end{equation}
which can also be seen from Equation~\ref{primaryDiffusionEquation}.

The \alfven-wave equation does not require any additional work beyond
that presented in \citet{BreitschwerdtEtAl1991}, except the
consideration of the loss term.  We start with the full \alfven-wave
equation:
\begin{eqnarray}
\frac{\partial}{\partial t} \left( \frac{(\delta B)^2}{4 \pi} \right)
+ \nabla \left( \frac{(\delta B)^2}{4 \pi} \left[\frac{3}{2}\bf{u} +
{\bf v}_{\rm A} \right] \right) = {\bf u} \nabla P_{\rm w} - {\bf v}_{\rm A} \nabla P_{\rm cr} + \mathcal{L},
\end{eqnarray}
where $\mathcal{L}$ represents other energy-gain, or, if negative,
energy-loss mechanisms for waves.  Simplifying this equation under the
assumptions of a steady-state, 1D system, we find:
\begin{eqnarray}
\frac{dP_w}{dz} & = & \frac{3u + \vA}{2(u + \vA)} \frac{P_w}{\rho}
\frac{d\rho}{dz} - \frac{\vA}{2(u+\vA)} \frac{dP_{cr}}{dz}  - 
\frac{3q}{2(u+\vA)} \frac{P_w}{\rho} +  \frac{\mathcal{L}}{2(u+\vA)}.
\end{eqnarray}

In our limit of $u \equiv 0$, $q \equiv 0$, we find:
\begin{equation}
\frac{dP_w}{dz} = \frac{1}{2} \frac{P_w}{\rho}
\frac{d\rho}{dz} - \frac{1}{2} \frac{dP_{cr}}{dz}   
 + \frac{\mathcal{L}}{2\vA}. \label{dPwEquation}
\end{equation}
In the absence of cosmic rays or wave-energy gains or losses,
Equation~\ref{dPwEquation} predicts $P_{\rm w} \propto \rho^{1/2}$, in
agreement with \citet{McKeeZweibel1995}.

We then substitute this equation for $dP_w/dz$ into
Equation~\ref{crPressureEq}.  Next, we need to define the
wave-pressure loss term, $\mathcal{L}$.

\subsection{\alfven Wave Loss Mechanisms}

This loss term, $\mathcal{L}$, represents the \alfven-wave damping
mechanisms; we consider both non-linear Landau damping (in the hot,
intercloud medium) as derived by \citet{Kulsrud2005}:
\begin{equation}
\dot{\mathcal{L}}_{\rm NLLD} = - 2 \cdot P_w \cdot \frac{3}{4} \sqrt{\frac{\pi}{2}}
\left( \frac{\delta B}{B} \right)^2 \frac{v_i}{c} \omega_{c,i} \label{landauDampingEq}
\end{equation}
(where we have substituted $\omega_{\rm Alfven} = k\vA$ with $k \sim
\frac{1}{r_{\rm g}} \sim \frac{c}{3}\frac{1}{\omega_{\rm c,i}}$), and
ion-neutral friction \citep[`case a' from][]{dePontieuEtAl2001}:
\begin{equation}
\dot{\mathcal{L}}_{\rm INF} = - 2 \cdot P_w \cdot 5 \times 10^{-15}
\sqrt{2} v_i \frac{\rho_n}{m_p}
\label{ionNeutralDampingEq}
\end{equation}
where $v_i$ is the ion thermal speed (which we set to $\sqrt{8 k_B
T/(\pi m_P)}$ throughout this work), $\delta B$ is the amplitude of
the \alfven-wave perturbation to the large-scale magnetic field ($B$),
and $\rho_n$ is the mass density of neutral atoms.  This equation
applies for ion-neutral friction between protons and hydrogen atoms;
lower cross-sections are necessary for the collision of carbon ions
with hydrogen atoms, for instance, in much cooler clouds.

\subsection{Energy Loss Mechanisms for Cosmic-Ray Protons}\label{lossFormula}

Cosmic-ray protons themselves are subject to energy loses due to pion
production, ionization, and Coulomb collisions
\citep[e.g.,][]{SkillingStrong1976}; these losses are required in
Equation~\ref{crPressureEq}, and are represented there with the term
$Q$.  To model these effects as a function of cosmic-ray-proton
energy, we utilize a modified version of the the energy loss function
given in \citet{Schlickeiser2002} as his Equation (5.3.58):
\begin{eqnarray}
-\left( \frac{dT}{dt} \right)_{\rm proton} (\beta) & = & 1.82 \times
 10^{-7} \times \bigg[ \left. (n_{HII} + n_{HI} + 2n_{H_2}) \times \left[
 5.44 (\gamma-1)^{7.64} \gamma^{0.75} H(1.75 - \gamma) H(\gamma - 1.43)
 \right. \right. \nonumber 
\\ 
& & \hspace{6.5cm}\left. \left. +~0.72 (\gamma-1)^{0.53} \gamma^{0.75} H(\gamma - 1.75)
 \right] \right. \nonumber \\
& & \hspace{3cm} +   1.69 n_e \frac{\beta^2}{\beta^3 +
 2.34\times10^{-5}(T_e/2 \times 10^6 K)^{3/2}} \nonumber \\
& & \hspace{4.5cm}\times H[\beta - 7.4 \times
 10^{-4}(T_e/2 \times 10^6 K)^{1/2}]  \nonumber \\
& & \hspace{3cm} + (n_{HI} + 2n_{H_2}) \frac{2 \beta^2}{10^{-6} + 2 \beta^3}
\times[1 + 0.0185
 (\ln \beta) H[\beta - 0.01]] \bigg]\, {\rm eV}\,{\rm s}^{-1}\label{protonLossEqn}
\end{eqnarray}
In this equation, $T$ is the kinetic energy per proton, $H[...]$ is
the Heaviside step function, $x$ is the ionization fraction, and
$\beta \equiv v/c$.  In this expression, the first term describes pion
production, the second gives Coulomb losses (normally negligible), and
the third term characterizes losses due to ionization.  The exact form
of this equation in \citet{Schlickeiser2002} made some assumptions
valid in the high-energy regime, but which could be problematic at
lower energies, so Equation~\ref{protonLossEqn} is our modified
version of the result from \citet{Schlickeiser2002}.  In particular,
the derivation of the pion-production loss term in
\citet{Schlickeiser2002} assumed $\gamma \gg 1$, and is undefined
below $\gamma = 1.3$, but we are interested in the regime $\gamma \sim
2$.  Equation~\ref{protonLossEqn} also more explicitly includes the
dependencies on $n_{HI}$, $n_{HII}$ and $n_{H_2}$ from earlier
expressions in \citet{Schlickeiser2002}.

Because we are especially interested in the cosmic-ray spectrum near
1\,\GeV, where the majority of cosmic-ray momentum is, we will start
by considering only the energy-loss rate near 1\,\GeV, but will
examine the impact of the variation in energy-loss rate in
Section~\ref{SensitivityToInitialConditions}.  This energy loss rate
gives the expression for $Q$ in the equation for the cosmic-ray
pressure (Eq.~\ref{crPressureEq}).

\section{Calculating the Cosmic-Ray Pressure in Clouds}\label{cloudModels}

\subsection{Boundary Conditions and Cloud Setup}\label{boundaryConditions}

To investigate cosmic rays penetrating diffuse clouds using
Equations~\ref{crPressureEq} and \ref{dPwEquation} (with definitions
from Eqs.~\ref{wavePressureEq}, \ref{BPressureEq},
\ref{kappaCRDefinition}, \ref{landauDampingEq},
\ref{ionNeutralDampingEq}, and \ref{protonLossEqn}), we set up a
simple interface between a highly-ionized plasma and a cool, neutral
cloud.

Our first use of this setup will be to check that the code reproduces
an analytical result: if there is no cosmic-ray diffusivity (the
cosmic rays are perfectly locked to \alfven waves), the cosmic-ray
pressure and \alfven speed change together to conserve the adiabatic
constant $P_{\rm cr} v_{\rm A}^{\gamma_{\rm cr}}$
\citep{BreitschwerdtEtAl1991}.  To test this, we define the density
contrast between the ionized and neutral phases with a simple {\it
  tanh} profile (the same profile we will use for all of our
calculations in this paper):
\begin{equation}
\rho(z) = \rho_{\rm init} + (\rho_{\rm final} - \rho_{\rm init}) \cdot
\frac{1}{2} \left( 1 + \tanh \left[  \frac{z - z_{\rm
      cloud~edge}}{\Delta z_{\rm edge}} \right] \right),
\end{equation}
with the center of the $\tanh$ function at $z_{\rm
  cloud~edge}=0.5$\,pc, a width of $\Delta z_{\rm edge}=0.05$\,pc; the
initial and final densities are set to (for this initial test)
$\rho_{\rm init} = 1$\,$m_{\rm p}$\,cm$^{-3}$ and $\rho_{\rm final} =
10$\,$m_{\rm p}$\,cm$^{-3}$.  We also set the ionization fraction to
$1$ throughout, and have removed all loss terms for the cosmic rays.
It is not possible to simply set $\kappa_{\rm cr} \equiv 0$, so we
integrate only Equation~\ref{crPressureEqSimplified} for $dP_{\rm
  cr}/dz$.  We then initialize the cosmic-ray pressure to
$10^{-12}$\,ergs\,cm$^{-3}$, and integrate the equations for
cosmic-ray and wave pressure into the cloud.  The resultant adiabatic
constant $P_{\rm cr} v_{\rm A}^{\gamma}$ shows only very small
variations; it is constant to 1 part in over $10^4$.  This verifies
the basic structure of the code.

\begin{figure}[h!]
\begin{center}
\includegraphics[angle=-90,width=12cm]{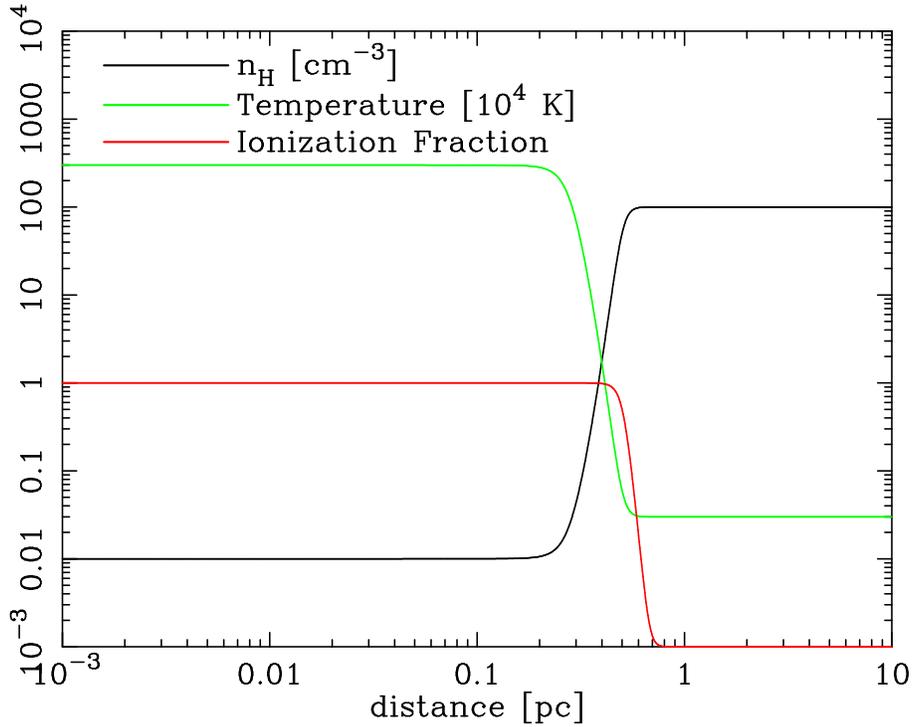}
\end{center}
\caption{The variations of gas density (in units of cm$^{-3}$),
  temperature (in units of $10^4$\,K), and ionization fraction as
  cosmic rays transition from the low-density, high-temperature ISM
  ($x_{\rm init} = 1$) to a simple cool cloud ($x_{\rm final} =
  10^{-3}$).  This transition is not meant to mimic any particular
  cloud, but is just a fiducial system to examine the change in
  cosmic-ray pressure.\label{cloudSetup}}
\end{figure}

Next, we define the complete transition from an ionized medium to a
cool cloud; this is the basic setup we will explore in this paper.
The ionized medium is defined to have $T = 3\times10^6$\,K, $\rho =
10^{-2}$\,$m_{\rm p}$\,cm$^{-3}$, $P_{\rm cr} =
10^{-12}$\,ergs\,cm$^{-3}$, $B = 3$\,$\mu$G (and the same value within
the cloud), and $x$ (the ionization fraction) set to unity.  As in the
previous test, the density is then defined to increase as a simple
{\it tanh} function; the final cloud density is set at
$\rho=100$\,$m_{\rm p}$\,cm$^{-3}$.  The ionization fraction, $x$, is
set to decrease as:
\begin{equation}
x(z) = x_{\rm final} + (x_{\rm init} - x_{\rm final}) \cdot
\frac{1}{2} \left( 1 + 
\tanh \left[ \frac{-(z -
    z_{\rm cloud~edge})}{\Delta z_{\rm edge}}\right] \right). \label{ionFracEqn}
\end{equation}
In this way, the ionization fraction is set to drop from a maximum
($x_{\rm init}$ = 1) to a minimum ($x_{\rm final} = 10^{-3}$) in a
very similar way to the drop in temperature, but lagging slightly
behind.  Meanwhile, we set the temperature to vary inversely as the
density to maintain a constant pressure.  The variation in density,
ionization fraction and temperature are all shown in
Figure~\ref{cloudSetup}.

We set the initial wave pressure such that the initial $\Pw$ is in
equilibrium with the initial cosmic-ray pressure gradient and the
initial wave damping.  For our fiducial setup, with a cosmic-ray
pressure gradient $dP_{\rm cr}/dz = -10^{-34}$\,ergs\,cm$^{-4}$
\citep[this value is the approximate cosmic-ray gradient in galactic
wind models in][]{EverettEtAl2010}, the equilibrium $\Pw$ is
approximately $2.3 \times 10^{-18}$\,ergs\,cm$^{-3}$, yielding an
equilibrium cosmic-ray diffusion coefficient of $1.5 \times
10^{27}$\,cm$^{2}$\,s$^{-1}$ in the ionized medium (or, put in more
transparent units, $\kappa_{\rm c} = 0.005$\,kpc$^2$/Myr).
Interestingly, this diffusion coefficient is in the range already
considered by some parameterized cosmic-ray diffusion models
\citep[e.g.,][]{CasanovaEtAl2011}.  However, this value is
approximately a factor of 20 to 50 lower than the overall Galactic
diffusion coefficient found from fitting diffusion models to
cosmic-ray spectra and gamma-ray emission, e.g.,
\citet{BerezinskiiEtAl1990}, \citet{PtuskinEtAl2008}, and
\citet{StrongEtAl2010}.  This difference can be understood as a result
of our choice of the input cosmic-ray pressure gradient; a smaller
pressure gradient (which could be quite reasonable for $P_{\rm cr}$
differences within the ISM) would lead to a smaller wave pressure, and
hence larger diffusivity; in order to explain the difference we see,
the cosmic-ray pressure gradient would have to be 2500 times smaller
in the ISM, or $dP_{\rm cr}/dz = -4\times10^{-38}$\,ergs\,cm$^{-4}$.
As we report in Section~\ref{SensitivityToInitialConditions}
(including Table~\ref{variationsInParameters}), even decreasing the
cosmic-ray pressure gradient down to $10^{-39}$\,ergs\,cm$^{-4}$
(corresponding to $\kappa_{\rm cr} \sim 5 \times
10^{29}$cm$^{2}$\,s$^{-1}$ or $1.7$\,kpc$^{2}$\,Myr$^{-1}$) does not
significantly change the results of our fiducial model.

We integrate the diffusion equation for cosmic-ray pressure until the
diffusion coefficient (calculated from Eq.~\ref{kappaCRDefinition})
exceeds $c L_{\rm cloud}$ (where $L_{\rm cloud}$ is the scale of the
cloud, here set to $10$\,pc), or until $P_{\rm cr}$ drops to zero,
which occurs in some extreme cases, as we discuss in
Section~\ref{SensitivityToInitialConditions}.  Deeper into the cloud,
beyond either of these limits, the cosmic rays are no longer
interacting with waves in the cloud (or in the case where $P_{\rm cr}$
drops to near zero, they are not generating waves in the cloud), and
so would enter a purely kinetic regime where we cannot treat them
hydrodynamically.  Effectively, this happens where-ever ion-neutral
damping takes over, damping the cosmic-ray generated waves, and
releasing the cosmic rays to free-stream into the cloud.

\subsection{Cosmic-Ray Pressure in the Cloud}\label{crPressureInCloud}

\begin{figure}[h!]
\begin{center}
\includegraphics[angle=-90,width=12cm]{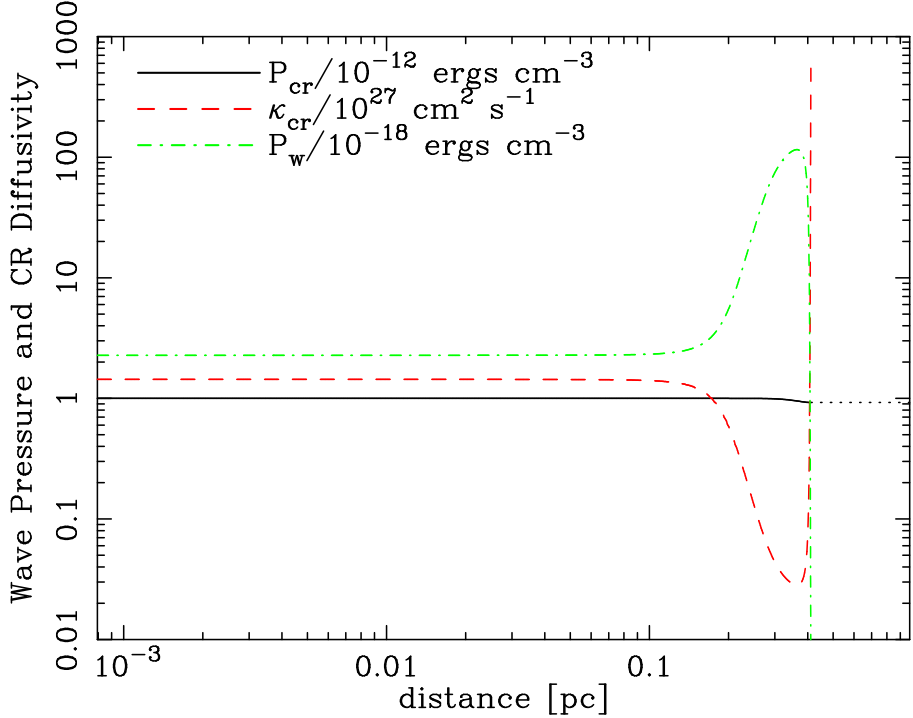}
\end{center}
\caption{The normalized \alfven-wave pressure ($\Pw$), cosmic-ray
  diffusivity ($\kappa_c$), and cosmic-ray pressure ($\Pcr$) as a
  function of distance from the external ionized medium (on the left)
  to the cool cloud (on the right).  $\Pw$ is initially in equilibrium
  outside the cloud, then increases on the cloud boundary as the
  density increases and as the cosmic-ray pressure drops slightly (by
  of order 5\%).  Near the peak of $\Pw$, where the growth slows,
  non-linear Landau damping starts to limits the growth of the waves;
  the destruction of $\Pw$ at $z = 0.4$\,pc is completely due to the
  onset of ion-neutral damping.  $\kappa_{\rm cr} \propto \Pw^{-1}$,
  so closely follows the development of the \alfven-wave
  pressure. Beyond $z = 0.4$\,pc, $\kappa_{\rm cr}$ exceeds $c L_{\rm
  cloud}$, and so the cosmic rays transition from the hydrodynamic
  approximation to the kinetic limit; beyond that point,
  the density of cosmic rays is constant (this is indicated by the
  dotted line).  Our treatment cannot describe cosmic rays in that
  limit, and so we stop tracking cosmic rays past that point.
  \label{crsInCloud} }
\end{figure}

We show the results for the cosmic-ray pressure, cosmic-ray
diffusivity, and \alfven-wave pressure in our fiducial model in
Figure~\ref{crsInCloud}.  This figure summarizes many of the key
findings of the paper: we see a slight ($7.5\%$) decrease in the
cosmic-ray pressure, with the coupling of cosmic rays to the gas
stopping at a distance of $\sim 0.4$\,pc into the cloud: further
inside the cloud, the cosmic rays decouple from the gas (this is
indicated by the dotted line in Fig.~\ref{crsInCloud}).  Beyond that
point of decoupling, the cosmic-ray pressure still formally exists
within the cloud (the cosmic rays still have a well-defined and
positive energy density), but that pressure is not communicated to the
gas in the cloud, so there is no cosmic-ray pressure force on the
cloud.  Interestingly, though, on the outskirts of the cloud, we see
that the cosmic-ray diffusivity actually decreases strongly because of
a sharp increase in the wave pressure.

\begin{figure}[h!]
\begin{center}
\includegraphics[angle=-90,width=12cm]{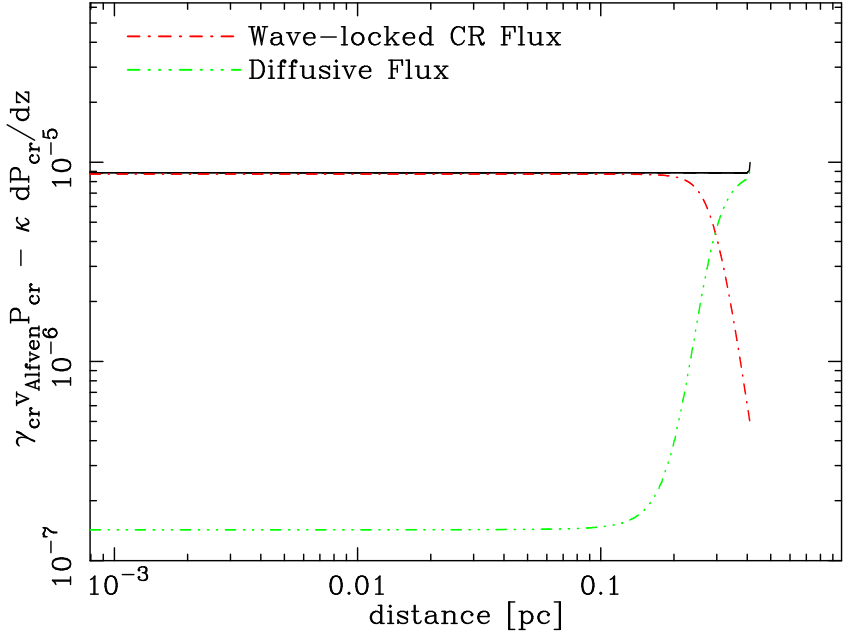}
\end{center}
\caption{The transition from an advectively-dominated cosmic-ray flux
  to the diffusively-dominated flux.  In summary, as the cosmic rays
  enter the cloud, the decrease in the \alfven speed leads to a sharp
  drop in the advective flux which must be compensated (in the steady
  state) by an increase in the diffusive flux.  Given the strong
  decrease in diffusivity, the cosmic-ray pressure is forced to drop
  in the cloud to yield the dominant diffusive flux.  The slight
  up-tick in the diffusive flux at the very end of the above traces (at
  0.4\,\pc) signals the face rise in diffusivity as ion-neutral damping
  becomes dominant, \alfven waves die off, and the cosmic rays start
  free-streaming through the cloud. \label{crFluxComparison}}
\end{figure}

How can we understand these effects intuitively?  The overall drop in
cosmic-ray coupling to near zero is easy to explain as the impact of
ion-neutral damping that destroys the \alfven waves, leading to
extremely weak coupling between the cosmic rays and the gas.  But why
the small decrease in the cosmic-ray pressure beforehand?  And why the
decrease in the diffusivity on the cloud boundary?  For these latter
effects, the key is to recognize that the solution is the only
self-consistent one that allows the initial cosmic-ray flux to be
transported through the cloud without a buildup in cosmic-ray pressure
at any point; this will be explained in more detail, below, but can be
seen by comparing the advective flux of cosmic rays (the cosmic-ray
flux in the ionized medium, where cosmic rays are largely tied to
\alfven waves) to the diffusive flux, as shown in
Figure~\ref{crFluxComparison}.

The incoming advective flux (plotted with the dot-dashed line in
Fig.~\ref{crFluxComparison}; this flux is given by the advection term
in Eq.~\ref{primaryDiffusionEquation}, and discussed further after
Eq.~\ref{simpleTransportEquation}, below) is set by the \alfven speed
and cosmic-ray pressure (the flux is $\gamma_{\rm cr} \vA \Pcr$).  As
the density increases into the cloud, the \alfven speed decreases, and
so the cosmic-ray flux that can be transported by the waves also
decreases, as shown in the downward trend of the dot-dashed line in
Figure~\ref{crFluxComparison}.  The only way to avoid a buildup of
cosmic-ray pressure on the outskirts of the cloud is to transfer that
flux of cosmic rays to a diffusive flux, $\kappa_{\rm cr} d\Pcr/dz$,
which depends on the cosmic-ray pressure gradient.  At the same time,
as an added effect, the increase in gas density leads to an increase
in \alfven-wave pressure \citep[see Eq.~\ref{dPwEquation}; see
also][]{McKeeZweibel1995}, and so the diffusion coefficient itself
must decrease, requiring a compensating change in the cosmic-ray
pressure, $P_{\rm cr}$.  The self-consistent solution is to therefore
have a larger cosmic-ray pressure gradient, which results in a
dominant diffusive flux into the cloud.

With the above general picture outlined, we return to the equation
governing cosmic-ray transport to explain the result more fully.  It
is easiest to look at a slightly expanded and simplified version of
the cosmic-ray transport equation (Equation~\ref{origCRTransportEq}).
If we write out that equation with the substitutions
\begin{eqnarray}
\frac{d\kappa_{\rm cr}}{dz} & = & - \frac{\kappa_{\rm cr}}{P_{\rm w}}
\frac{dP_{\rm w}}{dz} \\ \nonumber
\frac{d\vA}{dz} & = & -\frac{\vA}{2 \rho}\frac{d\rho}{dz},
\end{eqnarray}
we get:
\begin{eqnarray}
\gamma_{\rm cr} \left( v_{\rm A} \frac{dP_{\rm cr}}{dz} \right. & + &
\left.  \frac{dv_{\rm A}}{dz} P_{\rm cr} \right) + \left(
-\frac{d\kappa_{\rm cr}}{dz}
\frac{dP_{\rm cr}}{dz} - \kappa_{\rm cr} \frac{d^2 P_{\rm cr}}{dz^2}
\right) = (\gamma_{\rm cr} - 1) \left( v_{\rm A} \frac{dP_{\rm cr}}{dz} +
Q \right). \label{simpleTransportEquation}
\end{eqnarray}
The first term in parentheses yields the change in the advective flux,
while the second term in parenthesis gives the change in the diffusive
flux.  The terms on the right side of the equation give the various
energy-loss mechanisms: the cosmic-ray pressure decreases due to wave
generation, and also because of Coulomb interactions, pion productions,
or ionization (these latter three processes symbolized by $Q$).

This equation helps one to see why the cosmic-ray pressure changes
within clouds.  First, we consider the term that describes the changes
in advective flux.  Since $d\Pcr/dz < 0$, and $dv_{\rm A}/dz < 0$ in
the cloud (the \alfven speed decreases as the density increases), both
of these terms describe a \textit{drop in the advective flux of
cosmic-ray pressure into the cloud}.

Next, we consider the term that describes the change in diffusive
flux.  The first term in the second set of parentheses shows that the
diffusive flux drops as the diffusivity drops (while $d\Pcr/dz$ is
negative).  Since $d\kappa_{\rm cr}/dz = - \frac{\kappa_{cr}}{P_{\rm
    w}} \frac{dP_{\rm w}}{dz}$, as the wave pressure increases into
the cloud (because of the increase in density; see
Eq.~\ref{dPwEquation}), the diffusivity will decrease, and like the
advective flux term, this component of the diffusive flux also drops.
The second term then describes how the diffusive flux increases as
$d^2\Pcr/dz^2$ becomes more negative.

As the cosmic-ray flux progresses into the cloud, the only
self-consistent solution for the cosmic-ray pressure, in the
steady-state, is for $d^2\Pcr/dz^2$ to become increasingly negative so
that the cosmic-ray flux can become progressively diffusive and
continue to propagate into the cloud.  This explains why the
cosmic-ray pressure starts to decrease into the cloud.  This continues
until the ion-neutral damping destroys the waves, or $P_{\rm cr}$
falls to zero (in extreme cases).  \textit{Beyond this point, cosmic
  rays are no longer locked to \alfven waves when moving through the
  cloud, so no momentum or energy is communicated to the cool gas in
  the cloud.}  Further, beyond this point, the cosmic-ray density in
the cloud will be constant (neglecting cosmic-ray losses due to pion
production, Coulomb interactions, or ionization).  This result is
consistent with previous results concerning cosmic-ray density within
clouds \citep[e.g.,][]{Skilling1971}, although it conflicts with
recent studies \citep{PadoanScalo2005}, and addresses the cosmic-ray
pressure in the cloud, which has not previously been explored.

\subsection{Impact of Waves on the Cloud}\label{impactOfWaves}

\begin{figure}[h!]
\begin{center}
\includegraphics[angle=-90,width=12cm]{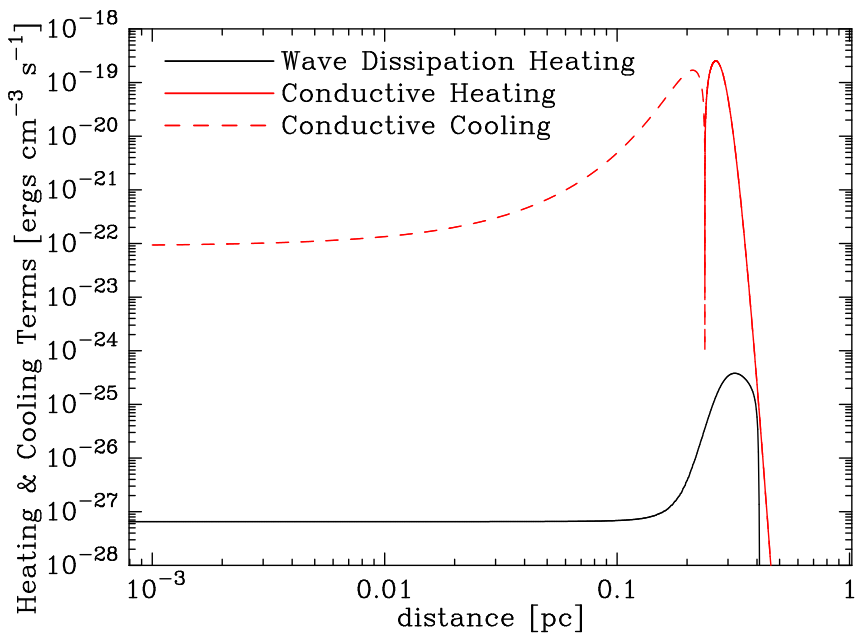}
\end{center}
\caption{Comparison of the cosmic-ray derived wave-dissipation heating
  (given in the black solid line) to the impact of thermal conduction
  between the cool cloud and the ambient hot plasma, shown in red.
  Thermal conduction heating dominates by six orders of magnitude in
  the cool-cloud interface, and thermal conduction also more than
  compensates for the heating due to cosmic rays in the ionized plasma
  surrounding the cloud, as shown with the dashed-red line.
  \label{conductiveHeatingComparison}}
\end{figure}

The relatively strong cosmic-ray pressure gradient and the relatively
high wave pressure on the outskirts of the cloud lead to heating on
the cloud boundary as \alfven waves are damped.
Given the magnetic-field configuration that we have assumed (a
constant magnetic field that directly connects the ambient medium to
the cloud), we can compare the heating from wave damping to that from
thermal conduction.  Our calculations show that thermal conduction is
dominant, as shown in Figure~\ref{conductiveHeatingComparison}.  To
estimate the conductive heating and cooling, we use the
thermal-conduction formulae of \citet{Pittard2007}: these estimates,
for the fiducial model that we use here, show that the maximum
conductive heating of the cool-cloud interface overpowers the maximum
wave-damping by six orders of magnitude, although there are small
regions where wave damping becomes comparable.  This shows that
cosmic-ray induced wave heating on the outskirts of the clouds is
largely insignificant compared to the effects of thermal conduction.
This is true even if we invoke a cooler ambient medium: if we modify
the parameters of the ionized gas to keep the same pressure, setting
$T = 10^4$\,K and $n = 3$\,cm$^{-3}$, the \alfven speed drops, so that
the overall cosmic-ray flux is lower.  In this regime, there is no
rise in the cosmic-ray wave pressure on the outskirts of the cloud
because of the relatively smaller rise in density, compared to the
fiducial model.  Therefore, there is less wave heating and the
conductive heating still dominates in the cloud, even for a cooler
ambient medium.

In regions where wave heating can grow to be competitive with
conductive heating, do the \alfven waves themselves become dynamically
important?  And do any of our test cases yield $P_{\rm w}/P_{\rm B}
\ga 1$, such that the wave growth becomes non-linear?  At our highest
magnetic-field strength, and at our highest cosmic-ray density, and in
our cases of maximum wave heating, the maximum wave pressure is
approximately 10\% of the ambient thermal pressure; in our fiducial
case, the wave pressure is only $\sim 10^{-4}$ of the thermal
pressure.  This shows that the wave pressure is not dynamically
important, but may become important at larger cosmic-ray fluxes.
Next, turning to the question of possible non-linearity in the wave
growth, the maximum $P_{\rm w}/P_{\rm B}$ that we find in our models
is $\sim 0.1$ (and for our fiducial model, $P_{\rm w}/P_{\rm B} \sim 3
\times 10^{-4}$ at its peak); this shows that wave growth is still in
the linear regime, although nearing the non-linear regime where
$P_{\rm w}/P_{\rm B} \sim 1$. So, we do not see significant wave
pressure or wave growth into the non-linear regime, although both of
these concerns merit careful monitoring as cosmic-ray fluxes increase
beyond those we consider.

\subsection{Sensitivity to Initial Conditions \&
  Assumptions}\label{SensitivityToInitialConditions}

Throughout this work, we have assumed a particular form for the hot
plasma/cloud interface, and a variety of parameters that describe both
the wind and cloud structure.  We have also specified boundary values
for $d\Pcr/dz$ and $\Pcr$.  We check for the limits of applicability
of our fiducial model, which showed that: (1) the cosmic-ray density
does not change significantly within the cloud, (2) the cosmic rays
are not coupled to gas within the cloud, and (3) cosmic-ray induced
heating is insignificant compared to thermal conduction.  We
investigate changes in these results as we modify the cloud interface
width, the cloud density, ionization profile, magnetic field strength,
initial cosmic-ray pressure, initial cosmic-ray pressure gradient,
cosmic-ray energy per nucleon (previously set to 1\,\GeV) and total
cosmic-ray energy density.  The parameters that we vary, and the range
over which those parameters are varied, are summarized in
Table~\ref{variationsInParameters}.

\subsubsection{Changes to Cosmic-Ray Pressure in the Cloud?}\label{pressureChanges}

First, do changes to any of these parameters modify the fiducial-model
result that cosmic rays do not couple to gas in the clouds?  No.  In
all of our calculations, we see this transition in the boundary of the
cool cloud.  In all of our models, the cosmic-ray streaming
instability fails at the cloud boundary (due either to ion-neutral
damping or a severe drop in cosmic-ray pressure), and cosmic rays do
not transfer energy and momentum to the gas within the clouds.

We do find changes in the cosmic-ray pressure on the cloud boundary,
before the transition from the hydrodynamic limit to the kinetic limit
(we refer to the last cosmic-ray hydrodynamic pressure within the
wave-locked zone as the ``minimum cosmic-ray pressure force'').  None
of the models show an increase in the cosmic-ray pressure in the
cloud.  The cosmic-ray pressure decreases slightly on the cloud
boundary in the fiducial case, so one might expect that as we increase
the penetration of wave-locked cosmic rays into the cloud, a larger
drop in cosmic-ray pressure on the cloud boundary would be seen.  This
is true, although we always see a quick transition to the kinetic
regime.

\begin{figure}
\begin{center}
\includegraphics[angle=-90,width=12cm]{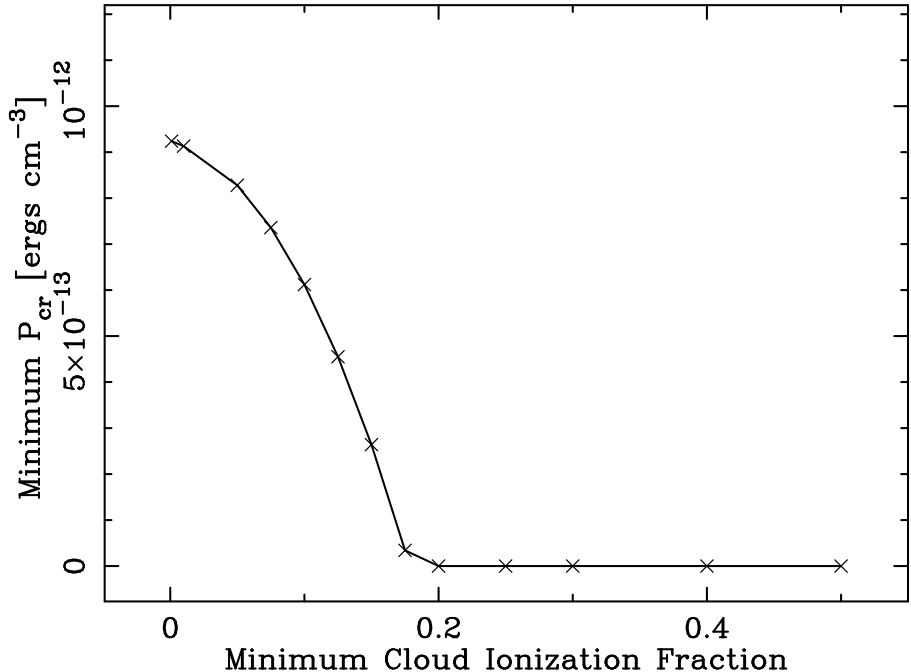}
\end{center}
\caption{The minimum cosmic-ray pressure force as a function of the
  lowest level of ionization in the cloud.
  \label{minPcrVsMinIonization}}
\end{figure}

\begin{figure}
\begin{center}
\includegraphics[angle=-90,width=12cm]{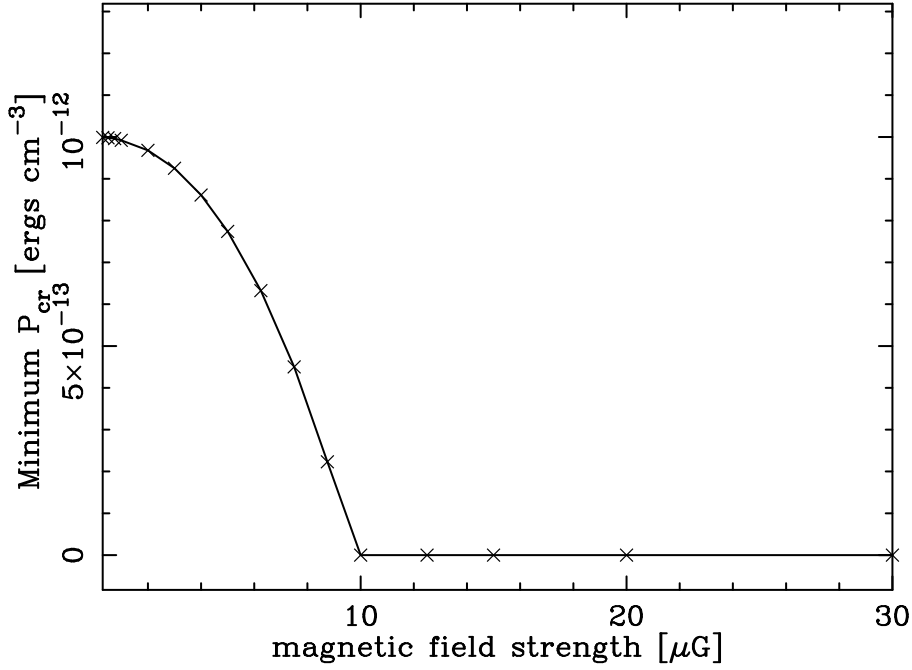}
\end{center}
\caption{The change in the minimum cosmic-ray pressure force with an
  increase in the (uniform) magnetic
  field. \label{magneticFieldChangePcr}}
\end{figure}

\begin{figure}[h!]
\begin{center}
\includegraphics[angle=-90,width=12cm]{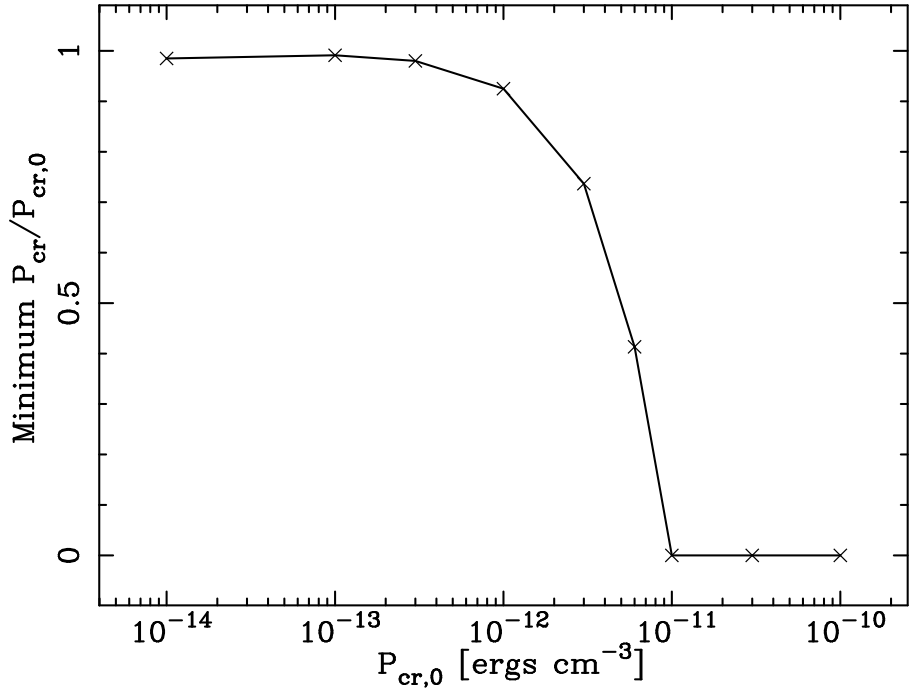}
\end{center}
\caption{The minimum cosmic-ray pressure force in the cloud as a
  function of the initial cosmic-ray pressure in the wind ($P_{\rm
    cr,0}$).  As the cosmic-ray pressure decreases, the effects of
  cosmic-ray pressure are felt to smaller and smaller depths into the
  cloud, with smaller and smaller differences between the external and
  internal cosmic-ray pressures force on the cloud boundary.
  \label{pcInitChangePcrRatio}}
\end{figure}

We find that the parameters that significantly affect the cosmic-ray
pressure on the boundary of the cloud are the minimum cloud ionization
fraction, the magnetic-field strength, and the initial cosmic-ray
pressure.  In the case of the minimum cloud ionization fraction, as
the cool cloud's minimum ionization level increases (from an
ionization fraction of 10$^{-3}$ to about 0.1-0.2), the onset of
ion-neutral damping is delayed, and the cosmic-ray pressure decreases
further, as \alfven waves penetrate further into the cloud, before
ion-neutral damping takes over.  Also, for increased magnetic field
strengths ($\sim 5-10$\muG), the advective cosmic-ray flux is higher
than in the fiducial case (because of a higher \alfven speed), and so
the cosmic-ray generated waves dominate farther into the cloud than in
the fiducial case before ion-neutral damping destroys the waves
altogether.  For $B > 10$\muG, $dP_{\rm cr}/dz$ is so strongly
negative that $P_{\rm cr}$ drops to zero.  The changes in the minimum
cosmic-ray pressure are shown in Figures~\ref{minPcrVsMinIonization}
and \ref{magneticFieldChangePcr}.  In a similar way, increasing the
initial cosmic-ray pressure also gives a stronger cosmic-ray flux at
the surface of the cloud, which penetrates further into the cloud
before ion-neutral damping takes over, and yields a change in the
cosmic-ray pressure at the cloud interface (see
Fig.~\ref{pcInitChangePcrRatio}).

\subsubsection{Changes to Boundary-Layer Heating in the Cloud?}

\begin{figure}
\begin{center}
\includegraphics[angle=-90,width=12cm]{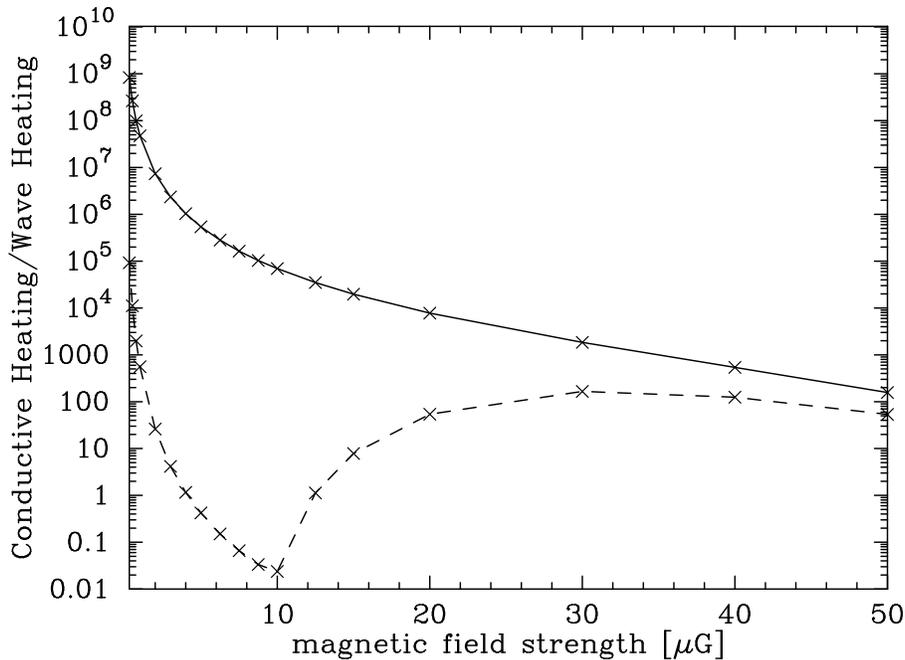}
\end{center}
\caption{The maximum and minimum ratio of conductive
  heating to wave-damping heating, plotted as a function of increase
  in the (uniform) magnetic field strength; the minimum ratio is shown
  with a dashed line, the maximum ratio is shown with a solid line.
  As the magnetic-field strength increases, the \alfven speed
  increases, and the flux of cosmic rays increases; this leads to a
  stronger wave pressure that extends deeper into the cloud, where,
  for intermediate field strengths of $4$\,\muG$ \la B \la 12$\,\muG,
  wave damping can compete with conductive heating.  Beyond that
  limit, the cosmic-ray pressure drops off too quickly into the cloud,
  and again, conductive heating dominates.
  \label{magneticFieldChangeHOverC}}
\end{figure}

\begin{figure}[h!]
\begin{center}
\includegraphics[angle=-90, width=12cm]{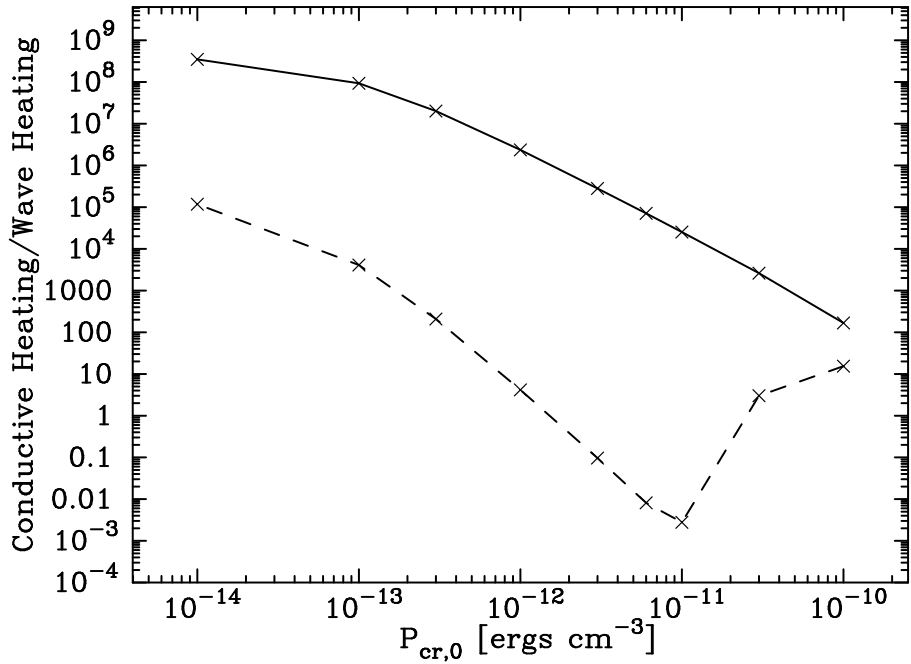}
\end{center}
\caption{As in Figure~\ref{magneticFieldChangeHOverC}, but
  as a function of increase in initial cosmic-ray pressure.  As with
  the variation of magnetic-field strength, there is a drop in the
  minimum ratio at intermediate cosmic-ray pressures ($1.5 \times
  10^{-12}$\,ergs\,cm$^{-3} \la P_{\rm cr,0} \la 2.5 \times
  10^{-11}$\,ergs\,cm$^{-3}$) due to the rise of the cosmic-ray flux,
  and then a drop at higher cosmic-ray pressures due to the strong
  drop-off in cosmic-ray pressure into the cloud.
  \label{pcInitChangeHOverC}}
\end{figure}

As shown in Section~\ref{impactOfWaves}, conductive heating overall
strongly dominates the heating from cosmic-ray generated \alfven
waves, although in very limited regimes, wave damping can compete with
conductive heating (see Fig.~\ref{conductiveHeatingComparison}).  As
we vary the parameters in Table~\ref{variationsInParameters}, the
heating at the cloud boundary changes, but in most cases and in most
places on the cloud boundary, those changes yield heating rates that
are orders of magnitude lower than that of thermal conduction.
However, we have found that, as we increase the magnetic-field
strength and the initial cosmic-ray pressure, there are small regions
of the cloud boundary that can become dominated by wave heating.  We
present these results in Figures~\ref{magneticFieldChangeHOverC} and
\ref{pcInitChangeHOverC}, which show the factors by which conductive
heating dominates the wave heating when we change the magnetic-field
strength and the initial cosmic-ray pressure, respectively.  Given the
varying strengths with distance of both the wave heating and
conductive heating (see Fig.~\ref{conductiveHeatingComparison}),
comparing these rates with a single ratio is too simplified, so we
compute the ratio of conductive to wave heating as a function of
distance, and plot both the maximum ratio of conductive heating to
wave heating (solid line), and the minimum ratio (dashed line).

As one might expect from our discussion of cosmic-ray pressure changes
in the previous section, when we increase the magnetic-field strength
or increase the external cosmic-ray pressure, the heating from
\alfven-wave damping increases (see
Fig.~\ref{magneticFieldChangeHOverC}), at least over a limited region
of parameters.  For some conditions ($1.5 \times
10^{-12}$\,ergs\,cm$^{-3} \la P_{\rm cr,0} \la 2.5 \times
10^{-11}$\,ergs\,cm$^{-3}$ and $4$\,\muG$ \la B \la 12$\,\muG), the
magnetic-field strength and cosmic-ray pressures are high enough to
yield a stronger cosmic-ray flux into the cloud (compared to our
fiducial case), which can give a large \alfven-wave amplitude in the
region where conductive heating drops off, and in those regions, the
wave heating can be greater than conductive heating.  Towards the
upper limit of our range of magnetic-field strength and cosmic-ray
pressure, however, the cosmic-ray pressure drops so strongly on the
boundary of the cloud (see Figs.~\ref{magneticFieldChangePcr} and
\ref{pcInitChangePcrRatio}) that the wave heating declines before it
can begin to compete with the drop-off in conductive heating.  Of
course, these results may vary with other cloud configurations, but
this analysis shows that it is possible for wave heating to compete
with conductive heating in limited circumstances.  It is also very
important to note, though, that as shown in
Figure~\ref{conductiveHeatingComparison}, conductive heating is still
dominant overall, despite the small regions in the cloud where wave
damping might be competitive.

\begin{deluxetable}{lll}
\tabletypesize{\footnotesize}
\tablenum{1}
\tablecaption{The parameters and values we have assumed in the fiducial
  study and the variations in those parameters that have been examined
  to test robustness of our results.} \label{variationsInParameters}
\tablehead{\colhead{Parameter} & \colhead{Fiducial Value} & \colhead{Range Tested}}
%\hline \hline
\startdata
Cloud Interface Width ($\Delta z_{\rm edge}$) & 0.05\,pc & 0.01\,pc -
0.1\,pc \\
Maximum Cloud Density ($\rho_{max}$) & 100\,cm$^{-3}$& 10\,cm$^{-3}$ -
1000\,cm$^{-3}$\\
Cloud Minimum Ionization ($x_{\rm ion}$) & 10$^{-3}$ & 0.01 - 0.5 \\
Magnetic Field Strength ($B$) & 3\muG & 0.3\muG - 30 \muG \\
Initial Cosmic-Ray Pressure ($P_{cr,0}$) & 10$^{-12}$\,ergs\,cm$^{-3}$
& 10$^{-14}$\,ergs\,cm$^{-3}$ - 10$^{-10}$\,ergs\,cm$^{-3}$ \\
Initial Cosmic-ray Pressure Gradient ($\frac{dP_{\rm cr}}{dz}|_0$) &
-10$^{-34}$\,ergs\,cm$^{-4}$ & -10$^{-31}$\,ergs\,cm$^{-4}$ -
-10$^{-39}$\,ergs\,cm$^{-4}$ \\
Cosmic Ray Energy per nucleon & 1\,\GeV & 1\,\MeV - 1\,\GeV
\enddata
\end{deluxetable}

\section{Limitations, Comparison to Previous Work, \& Implications}\label{conclusions}

This calculation makes a number of important assumptions, which we
summarize here.  First of all, we assume that there is a large-scale
magnetic field which threads both the ionized gas and the cool cloud
at constant strength.  If the field strength changes, magnetic
mirroring effects could also be important
\citep[e.g.,][]{CesarskyVoelk1978,Chandran2000a}, but we do not
consider such effects here.  This is a simplification, but one that is
observationally warranted: \citet{CrutcherEtAl2010} find that the
magnetic-field strength does not scale with density for $n \la
300$\,cm$^{-3}$ (below the densities that are important in this work),
and \citet{HeilesTroland2005} find that the magnetic-field strength in
the cold, neutral medium of the ISM is of the same order as in other
ISM components \citep[this may perhaps be understood by ambipolar
diffusion, see][]{Zweibel2002}.

Also, we have considered only the role of the streaming instability in
scattering cosmic rays; however, this is the dominant process for
$\sim 1$\,\GeV cosmic rays \citep{Kulsrud2005}.  It is important to
ask if other forms of turbulence may help scatter cosmic rays;
however, turbulent wave modes, especially on the very small scales of
the cosmic-ray gyroradius ($r_{\rm g} \sim 10^{12}$\,cm) are believed
to lose energy very quickly in molecular clouds
\citep[e.g.,][]{MacLowEtAl1998, OstrikerEtAl2001}.  In addition, those
modes do not scatter cosmic rays efficiently
\citep[e.g.,][]{Chandran2000b,YanLazarian2004}, so we do not expect
the diffusivity to be affected by those processes.

Finally, when considering the role of cosmic rays in galactic winds,
we do not consider the drop in velocity of the cool cloud as cosmic
rays enter the cloud.  Such a drop in velocity is likely to occur, as
the hot-wind component is usually inferred to move at thousands of
km/s \citep[e.g.,][]{StricklandHeckman2009}, and the clouds have
velocities of $\la 300-500$\,km/s \citep[e.g.,][]{Martin2005,
  SteidelEtAl2010}, so that the velocity difference could be of order
a factor of four.  In our calculations, though, the \alfven speed
already drops by an order of magnitude; a smaller drop in the bulk gas
velocity would have the same, but smaller, effect on the transition of
cosmic rays into the cool cloud.  We infer that such a change in the
bulk velocity will therefore only further increase the cosmic-ray flux
injected into the clouds, and will have a similar impact on cosmic-ray
pressure and heating as increases in the magnetic field strength and
cosmic-ray pressure have.  Therefore, the effects of increased
velocity in the hot, ambient medium would not qualitatively change the
results presented here.

How do our results compare to previous work?  In the two sections
below, we consider previous work on cosmic rays in cool clouds in the
interstellar medium and in the galactic winds.

\subsection{Cosmic Rays in the Multiphase ISM}

The small drop in cosmic-ray pressure that we predict might seem
similar to the ``exclusion'' of cosmic rays from clouds that was found
by both \citet{SkillingStrong1976} and \citet{CesarskyVoelk1978}.
However, our model differs from those previous works in significant
ways, and therefore this new model brings new effects to light.  In
those earlier papers, the exclusion of cosmic rays was driven by a
drop in cosmic-ray density at $E \la 300$\,\MeV due to ionization
loses; as low-energy cosmic rays traversed the entire cloud, they were
destroyed in collisions.  It is also important that those works
assumed a uniform cosmic-ray pressure outside of the cloud, and so
assumed that the streaming instability only operated on the outskirts
of the clouds.

In comparison, our work assumes an initial cosmic-ray pressure
gradient (see Section~\ref{boundaryConditions}), such that the
streaming instability operates throughout the hot phase of the ISM.
This assumption seems reasonable as we have found that even very small
cosmic-ray pressure gradients lead to the growth of the streaming
instability, which seems plausible since the streaming instability can
explain the observed near isotropy of cosmic rays at the Earth
\citep[see, e.g., ][]{Kulsrud2005}.  With this configuration, our
models show a qualitatively new effect at higher energies: the
transition of cosmic rays from the advective regime to the
diffusively-dominated regime leads to a (small) cosmic-ray pressure
drop in the outskirts of clouds.  This effect is also present at low
energies (down to $E = 1$\,\MeV), and is more important in our models
than ionization losses (at least for cosmic rays just entering the
outskirts of the clouds).

In the end, it is difficult to directly compare these models, but we
believe that the assumption of zero cosmic-ray pressure gradient in
the ISM (the assumption made in the papers of
\citeauthor{SkillingStrong1976} and \citeauthor{CesarskyVoelk1978}) is
not correct, and that therefore the ``exclusion'' of cosmic rays from
cool clouds is not as strong as calculated in those papers (as we find
in this work); in our models, we find only very small drops in
cosmic-ray pressure within clouds.  However, this cannot yet be a
clear-cut comparison: we stress that the impact of low-energy
cosmic-ray losses throughout the entire cloud (instead of just on the
boundary, as we calculate here) is indeed important for large cloud
columns as the previous papers have pointed out, and so should be
considered further.  However, this realization points to a prediction:
the observation of a \textit{strong} decrease in the density of cosmic
rays in cool clouds at low energy (by an order of magnitude or two at
$E \sim 10$\,\MeV) would argue for the picture of \citet[][see their
Fig.~1]{SkillingStrong1976} and \citet{CesarskyVoelk1978}, whereas a
relatively constant density of low-energy cosmic rays would argue for
the model in this paper.  Recent results that the cosmic-ray density
does not change much or may even be higher than expected
\citep[e.g.,][]{McCallEtAl2003, IndrioloEtAl2010b} seem to point
towards the model presented here, but more detailed studies are
needed, both in the theory and perhaps in further observations.
Overall, such studies would be useful as they may also allow insight
into the presence or absence of a cosmic-ray gradient in the ISM, and
therefore help us understand the propagation of cosmic rays in the
ISM.  (Of course, there are extremes of parameter space where the
current model also yields drops in the cosmic-ray pressure within
clouds, but these are rather extreme values; see
Section~\ref{SensitivityToInitialConditions}.)  We stress that both
papers agree, however, that the density of cosmic rays is relatively
constant in clouds at $E \sim 1$\,\GeV, which is the main area of
focus in this work, and that, of course, cosmic-ray ionization losses
will occur, and should be modeled in a complete picture of the full
cloud for the most accurate comparison.

%We stress again, however, that most of our work here concerns cosmic
%rays with energies near $1\,\GeV$, since cosmic rays of this energy
%dominate the cosmic-ray pressure.  This difference in energies is
%crucial: at the energies that we are interested in, cosmic-ray losses
%due to ionization and Coulomb interactions are not significant,
%especially on the boundary of the cloud.  Instead, the cosmic-ray
%pressure changes here are due to a transition to the
%diffusively-dominated regime, as explained in
%Section~\ref{crPressureInCloud}.  Also, the change in cosmic-ray
%density that we predict at $\sim 1\,\GeV$ is generally smaller than
%that predicted by either \citet{SkillingStrong1976} or
%\citet{CesarskyVoelk1978} at $\sim 100\,\MeV$;

Putting aside those concerns about low-energy cosmic rays, we conclude
that for most of the parameter space of the ISM, the cosmic-ray
pressure does not change significantly before the cosmic rays become
free-streaming.  One might ask if the decreased transit time of cosmic
rays that are not locked to \alfven waves (with transit times of order
$\sqrt{3} L_{\rm cloud}/c$ vs $L_{\rm cloud}/\vA$) might help decrease
the gamma-ray emission, but since cosmic rays can be scattered back
and forth across the cloud by \alfven waves on either side of the
cloud, it is not clear that the total residence time of cosmic rays
inside clouds is significantly different from those for cosmic rays
that are locked to \alfven waves in a similar volume.  Overall, at
least, these results suggest that for clouds illuminated by cosmic
rays and perhaps observed in gamma rays
\citep[e.g.,][]{GabiciEtAl2007, TorresEtAl2008, AbdoEtAl2010}, the
cosmic-ray density inferred within the clouds should not be
significantly different than the density inferred outside of the
clouds.  We only predict decreases for the cosmic-ray density in
clouds when the ambient medium has a very high cosmic-ray pressure (an
order of magnitude higher than in the solar neighborhood) or when the
(assumed uniform) magnetic-field strength is somewhat high (with
values approximately twice the local field strength), or is more
highly ionized than our fiducial case (minimum ionization fractions
greater than $\sim 0.1$ lead to significant drops in the cosmic-ray
density).

In contrast to these results, \citet{PadoanScalo2005} inferred that
cosmic-ray density would increase significantly within the cool
clouds; such an increase in cosmic-ray density was predicted for
clouds of density $\la 100$\percc.  We do not see this effect.  It
appears that the equations in \citet{PadoanScalo2005} assume that the
cosmic rays always travel at the local \alfven speed in cool clouds.
In contrast, the analysis presented in this paper shows that
diffusivity quickly becomes dominant, and does not allow such a strong
increase in cosmic-ray density.

\subsection{Cosmic Rays in Multiphase Galactic Winds}

As mentioned in the introduction, acceleration of cool clouds in winds
without those clouds being quickly destroyed (on timescales of order a
few times the sound crossing time in the external medium) is difficult
to understand in clouds embedded in galactic winds.  If cosmic-ray
pressure had a smooth gradient within the cloud, perhaps it could
input momentum to the cloud in a distributed and less disruptive way;
however, we find that cosmic rays do not couple to gas within the body
of the cloud.  There is therefore no ``body force'' on the whole cloud
due to cosmic rays; like ram pressure, cosmic rays would exert a
pressure difference on the boundary that could perhaps help accelerate
the cloud, but would add to the pressure differences already acting to
disrupt the cloud.  This may indicate another reason why radiative
acceleration of clouds might be important \citep[see,
e.g.,][]{MurrayEtAl2010}, although perhaps any mechanism that imparts
momentum to clouds will introduce instabilities \citep[in the case of
radiation pressure, see][]{Mathews1986}.  Connecting to our other main
results, perhaps the heating on the outskirts of clouds by wave
damping could help pressurize the surface of the cloud and slow down
other instabilities.  Meanwhile, an understanding of the mechanisms by
which cosmic-ray density can drop in clouds will be important for
predictions of gamma-ray emission.

\acknowledgements We thank Mitch Begelman, Andrey Beresnyak, Jungyeon
Cho, Alex Lazarian, and Chris Matzner for helpful questions and
conversations.  We also thank the referee for helpful comments on the
manuscript.  J.E.E. thanks the Canadian Institute for Theoretical
Astrophysics for their hospitality, where some of this work was
completed.  The work in this paper has made use of NASA's Astrophysics
Data System Bibliographic Services.  J.E.E. and E.G.Z. acknowledge the
support of NSF grants PHY 0821899, AST 0907837, and AST 0507367 \& PHY
0215581 (to the Center for Magnetic Self Organization in Laboratory
and Astrophysical Plasmas), as well as support from the NASA
\textit{Fermi} grant NNX10AO50G.

\bibliography{crsAndClouds}

\end{document}